\documentstyle[12pt]{article}
\newcommand{\PSbox}[3]{\mbox{\rule{0in}{#3}\includegraphics{#1}\hspace{#2}}}

\newcommand{\jref}[4]{{\it #1} {\bf #2}, #3 (#4)}

\newcommand{\NPB}[3]{\jref{Nucl.\ Phys.}{B#1}{#2}{#3}}

\newcommand{\PLB}[3]{\jref{Phys.\ Lett.}{#1B}{#2}{#3}}

\newcommand{\PRD}[3]{\jref{Phys.\ Rev.}{D#1}{#2}{#3}}

\newcommand{\PRL}[3]{\jref{Phys.\ Rev.\ Lett.}{#1}{#2}{#3}}

\newcommand{\PTP}[3]{\jref{Prog.\ Theor.\ Phys.}{#1}{#2}{#3}}

\newcommand{\nc}{\newcommand}
\nc{\beq}{\begin{equation}} \nc{\eeq}{\end{equation}}
\nc{\beqa}{\begin{eqnarray}} \nc{\eeqa}{\end{eqnarray}}
\nc{\lsim}{\begin{array}{c}\,\sim\vspace{-21pt}\\< \end{array}}
\nc{\gsim}{\begin{array}{c}\sim\vspace{-21pt}\\> \end{array}}

\begin{document}

\begin{titlepage}

\begin{center}

\vspace{2cm}

{\hbox to\hsize{hep-ph/9707386 \hfill  MIT-CTP-2654}}

 \vspace{2cm}

\bigskip

{\Large \bf  Composite Intermediary and Mediator Models
of Gauge-Mediated Supersymmetry Breaking\footnote{Work supported in part by 
the US Department of Energy  under cooperative
                 agreement \#DE-FC02-94ER40818.}}

\bigskip

 {\bf Csaba Cs\'aki,}\footnote{Address after August 1:
  Department of Physics,  University of California, Berkeley, CA 94720.}
{\bf Lisa Randall}\footnote{Supported in part by 
an NSF Young Investigator Award, 
an Alfred P. Sloan Foundation Fellowship and a  DOE Outstanding Junior 
Investigator Award.} {\bf and
Witold Skiba}\footnote{Address after September 1: Department of Physics, 
University of California at San Diego, La Jolla  CA 92093.} \\

 \bigskip

 { \small \it   Center for Theoretical Physics\\
    Laboratory of Nuclear Science and Department of Physics\\
    Massachusetts Institute of Technology\\
Cambridge, MA 02139, USA \\}

  {\tt  csaki@mit.edu, lisa@ctptop.mit.edu, skiba@mit.edu } 

 \bigskip

 \bigskip

\vspace{2cm}
          
{\bf Abstract}
\end{center} 

We discuss gauge-mediated models which employ a tree-level
mass term in the superpotential. We give
explicit composite realizations in which the mass terms
are not fundamental. Instead, they arise
as effective terms in the superpotential from
confining gauge dynamics.

   \end{titlepage}

\renewcommand{\thepage}{\arabic{page}}

\setcounter{page}{1}

\setcounter{equation}{0}

\baselineskip=18pt
 
\section{Introduction}

 Gauge-mediated supersymmetry breaking is in principle
an attractive alternative to gravity-mediated supersymmetry
breaking in that the flavor changing neutral current (FCNC)
problem is automatically solved. However many of the
existing models are complicated or contrived so it is worthwhile
to explore alternative model-building ideas.

In Ref.~\cite{me}, two classes of gauge-mediated models
were introduced. In one class, called ``Intermediary" models,
there are two massive singlets, one of which couples to the dynamical
supersymmetry breaking sector and the other one to the  messenger quarks, 
but for which
there is a Dirac mass term coupling the two. Upon integrating
out the singlet there is automatically a coupling between the
dynamical supersymmetry breaking (DSB)  and visible sectors which can 
communicate
supersymmetry breaking to the messenger quarks and hence
to the visible sector. The phenomenology of these models
is similar in many respects to other messenger models~\cite{DNNS,jmr2,we}.

In the other class of models, called ``Mediator" models,
there are massive mediator fields which carry the gauge charge of a gauged 
messenger
group of the DSB sector and also standard model gauge charge.
These fields therefore permit communication of supersymmetry
breaking, but at high loop order. The scalar partners obtain a two-loop
supersymmetry breaking mass, whereas the gauginos obtain
a three-loop mass. The phenomenology of these models is therefore
distinctive (and probably somewhat more fine-tuned) in that
the scalars are expected to be about an order of magnitude heavier
than the gauginos.

The advantage of both these models is that there is no complicated
superpotential required in order to communicate supersymmetry
breaking to a fundamental singlet which is coupled to the messenger
quarks. In the intermediary models, the communication is automatic
upon integrating out the singlet. In the mediator models, a dynamical
messenger sector \cite{pt,jmr,mur} which automatically communicates
supersymmetry breaking to messenger gauginos and ultimately
to the visible sector is assumed.  The relative simplicity of these
scenarios occurs because we have not made the very restrictive
assumption about the absence of mass terms in the superpotential.
When this assumption is relaxed, more direct communication of supersymmetry 
breaking in a stable or sufficiently stable
vacuum occurs reasonably simply.

However, fundamental mass terms with a scale other
than that determined by the Planck scale or non-perturbative
gauge dynamics would be a strong assumption. Our philosophy
in generating these models was to first see what works,
with the assumption that it would be relatively straightforward
to realize the required masses and couplings in a composite
scenario.

In this paper, we give explicit realizations of composite dynamics
which produce successful Intermediary and Mediator models.
These models serve as existence proofs and allow a more accurate
determination of how complicated these models are compared
to the microscopic realization of alternative gauge-mediated
scenarios. Of course it is conceivable that simpler composite implementations
exist; the models presented here set an upper bound for complexity.

In the second section we discuss composite intermediary
models. We give explicit realizations of the underlying
gauge dynamics. We also explain
the motivation for the various elements assumed
in these models.  In the third section we discuss composite
mediator scenarios. 

Alternative models based on a dynamical messenger sector generated 
by an inverted hierarchy have been recently presented in Ref.~\cite{mur}.

\section{Composite Intermediary Models}

The intermediary models employ two massive gauge singlet fields
$S$ and $\overline{S}$ to communicate supersymmetry breaking to the
visible sector. The role of these singlets is to generate
an effective non-renormalizable operator connecting the
fields in the supersymmetry breaking sector to the messenger
quarks $Q,\overline{Q}$. Such an operator is suppressed by the mass
of the singlet fields $M_S$. The effects of this operator will be
identical to the coupling of the gauge singlet to the
messenger quarks of the original model of DNNS~\cite{DNNS} without 
requiring a messenger gauge group or complicated superpotential.

In the simplest version of the intermediary models, 
one has a pair of
vector-like fields $V$ and $\overline{V}$ in the supersymmetry 
breaking sector which have a non-vanishing F-term. 
One of the singlets, $S$, couples to the vector-like flavor
of the supersymmetry breaking sector $V,\overline{V}$, while the
other singlet, $\overline{S}$, couples to the messenger quarks
$Q,\overline{Q}$. The two sectors communicate only via the 
mass term for the two singlets, $M_S S\overline{S}$. To see the
effect of this coupling we examine the superpotential
\begin{equation}
\label{interm}
  W=SV\overline{V}+\overline{S}Q\overline{Q}+M_S S\overline{S}.
\end{equation}
Integrating out the massive singlets $S,\overline{S}$  
produces an effective superpotential term
\begin{equation}
\label{effop}
  W_{eff}=-\frac{(V\overline{V})(Q\overline{Q})}{M_S},
\end{equation}
below the mass scale $M_S$
which can mimic the effects of the coupling of the singlet to the
messenger quarks in the original DNNS models~\cite{DNNS},
provided that the composite field $V\overline{V}$ has the correct
F-type expectation value.
In order to obtain the correct gluino and squark masses one might also require
an explicit mass term $M_Q Q\overline{Q}$ for the messenger quarks and a mass 
term $M_VV\overline{V}$. We will see examples where the term
$M_Q Q\overline{Q}$ is not necessary; the requirement on $M_V V\overline{V}$ 
depends on whether it is necessary to lift dangerous flat directions. It is
most convenient to assume it is present.
 Thus the full 
superpotential required for the intermediary models
is  of the form
\begin{equation}
\label{suppot}
W=SV\overline{V}+\overline{S}Q\overline{Q}+M_SS\overline{S}+
  M_QQ\overline{Q}+M_VV\overline{V}.
\end{equation}

Once we have the superpotential of Eq.~\ref{suppot}
with the right values of mass terms and the correct F-terms
for the fields $V\overline{V}$, the mechanism described above
generates the effective $-\frac{(V\overline{V})(Q\overline{Q})}{M_S}$
coupling and supersymmetry breaking is mediated to the
visible sector in the usual way. This mechanism is clearly 
rather simple and generic.

The question one has
to address however is how the superpotential of
Eq.~\ref{suppot} can be obtained without introducing
artificially small mass terms (compared to the Planck-scale)
while simultaneously including unsuppressed Yukawa 
couplings. We will present several models
in which the effects of confining dynamics are exactly
to produce the superpotential of Eq.~\ref{suppot}.
In these models the singlets $S,\overline{S}$, the messenger quarks
$Q,\overline{Q}$ and the vector-like flavor $V,\overline{V}$ are
composites of an underlying strongly interacting gauge
group. The effect of the confining dynamics will be to bind
the ``preon'' fields into the composites $S,\overline{S},Q,\overline{Q},
V,\overline{V}$ and to generate an ${\cal O}(1)$ Yukawa coupling for
these fields required for the superpotential of Eq.~\ref{suppot}.
The mass terms will be obtained from non-renormalizable 
preon operators that turn into mass terms for the composite
fields after confinement~\cite{dsb}. Throughout this paper
we will assume that the only scales present in the models 
are the dynamical scales of various
confining groups and and the Planck-scale suppressing
possible non-renormalizable operators. 

\subsection{The confining sector\label{comp}}

We will first present a model which generates the $M_Q Q \overline{Q}$ 
and $\lambda \overline{S} Q \overline{Q}$ terms, where $\lambda$ is of order
one. These terms are a part of the superpotential in Eq.~\ref{suppot}.
It will become clear that the method is general and can be used with
other gauge groups.

The idea is to use a confining theory whose global symmetries contain
an $SU(5)$ subgroup and then gauge this subgroup which we identify
with the ordinary $SU(5)$. The best known example of a confining
supersymmetric theory is SUSY QCD with the
number of flavors equal to the number of colors plus one. Here we
want to gauge an $SU(5)$ symmetry, so we can take an $SU(4)$ theory
with five flavors of fundamentals $q$ and antifundamentals $\overline{q}$.
These  fields transform as follows under the strong $SU(4)$ and
global symmetry $SU(5)\times SU(5)$.
\begin{displaymath}
  \begin{array}{c|c|cc||c}
    & SU(4) & SU(5) & SU(5) & SU(5)_D\\ \hline
    q & 4 & 5 & 1 & 5 \\
    \overline{q} & \overline{4} & 1 & \overline{5} & \overline{5}
  \end{array}
\end{displaymath}
The $SU(5)_D$ in the above table is not an additional symmetry,
but the diagonal subgroup of $SU(5)\times SU(5)$ which we gauge.

This theory confines at low energies~\cite{SeibergPRD}. The composite
spectrum consists of the mesons $M\sim q\overline{q}$, the
baryons $B\sim q^4$ and the antibaryons $\overline{B}\sim \overline{q}^4$,
whose transformation properties are 
\begin{displaymath}
  \begin{array}{c|cc||c}
    & SU(5) & SU(5) & SU(5)_D\\ \hline
    M\equiv \Sigma +\overline{S} & 5 & \overline{5} & 24 + 1 \\
    B\equiv Q& \overline{5}& 1 & \overline{5} \\
    \overline{B}\equiv \overline{Q} & 1 & 5 & 5
  \end{array}.
\end{displaymath}
Again, we denoted explicitly the gauged subgroup of $SU(5)\times SU(5)$
in the last column. The strong dynamics generates a superpotential
\begin{displaymath}
  W_{conf}=\frac{1}{\Lambda^7}({\rm det} M-BM\overline{B}),
\end{displaymath}
which needs to be expressed in terms of fields with definite $SU(5)_D$
quantum numbers. The composite meson field contains an adjoint and a singlet
$\overline{S}$ of the gauged $SU(5)$, while the baryon and the 
antibaryon are identified with the messenger quarks.
Among other terms, the above superpotential contains the term
$\overline{S} Q \overline{Q}$, with a coefficient of order one for canonically
normalized fields.

We also want to generate masses for the fields $Q$ and $\overline{Q}$. This 
can be achieved by adding a tree-level term in the microscopic theory
of the form $W_{tree}=\frac{q^4 \overline{q}^4}{M_P^5}$. For normalized fields
this translates to giving mass $M_Q=\frac{\Lambda_{SU(4)}^6}{M_P^5}$
for the composite fields $Q$, $\overline{Q}$. This way we have achieved
generating a mass term for $Q$ and $\overline{Q}$ and a Yukawa term
with the coupling of order one for the composite fields.
This mechanism will be useful in a large class of our examples.

It is now obvious that the same mechanism can be used for fields transforming
as fundamentals under an $SU(N)$ group. The confining interactions are then
based on an $SU(N-1)$ gauge group. As before, the Yukawa coupling between
the composite fundamentals and the composite singlet has a coefficient of
order one. The mass term, on the other hand, is obtained from a tree-level
superpotential term of the form $\frac{(q')^{N-1}
(\overline{q}')^{N-1}}{M_P^{2 N-5}}$. Therefore the mass of the fundamental
fields is $\frac{\Lambda_{SU(N-1)}^{2 N-4}}{M_P^{2N-5}}$. In this scenario,
the adjoint of the gauged subgroup has to be given a mass by adding
a tree-level term which becomes a mass term after confinement.

We can now generate the remaining terms in the superpotential of
Eq.~\ref{suppot} by the same method. Since in the first model presented
below $V$ and $\overline{V}$
transform under $SU(3)$, we take a confining $SU(2)$ theory and obtain
appropriate Yukawa coupling $S V \overline{V}$ and a mass
$\frac{\Lambda^2_{SU(2)}}{M_P} V \overline{V}$. We also can obtain a mass term
for $S$ and $\overline{S}$. This requires a tree-level term 
$\frac{(q \overline{q}) (q' \overline{q}')}{M_P}$, which gives
$M_S=\frac{\Lambda_{SU(2)} \Lambda_{SU(3)}}{M_P}$.
To summarize, assuming that the fields 
$V,\overline{V},Q,\overline{Q},S,\overline{S}$ are composites results in the 
superpotential of Eq.~\ref{suppot}, where the trilinear 
Yukawas are order one due to the confining dynamics,
while the masses are given by 
\begin{equation}
\label{masses}
  M_Q=\frac{\Lambda_{SU(4)}^6}{M_P^5},\; \; 
  M_V=\frac{\Lambda_{SU(2)}^2}{M_P},\; \;
  M_S=\frac{\Lambda_{SU(2)}\Lambda_{SU(4)}}{M_P}.
\end{equation}

\subsection{The 3-2 model as the supersymmetry breaking sector\label{32sec}}

In the first explicit 
example we use the ``3-2 model'' of Affleck, Dine
and Seiberg~\cite{ADS} for the dynamical supersymmetry
breaking sector, but add an additional flavor 
($V+\overline{V}$) which transforms as $3+\overline{3}$ under the 
$SU(3)$. This modified model clearly breaks supersymmetry
when the extra flavor is massive.
The superpotential terms $SV\overline{V}+M_VV\overline{V}$
are generated by a confining $SU(2)$ group with three 
flavors (six doublets) as described above, which will result
in a mass term $M_V=\frac{\Lambda_{SU(2)}^2}{M_P}$ and the
order one Yukawa coupling $SV\overline{V}$. The superpotential
terms for $\overline{S}$ and the messenger quarks are generated 
by a confining $SU(4)$ group as described above, resulting in
$M_Q=\frac{\Lambda_{SU(4)}^6}{M_P^5}$. Finally, the mass for the
singlets is given by $M_S=\frac{\Lambda_{SU(2)}
\Lambda_{SU(4)}}{M_P}$. The fields of the supersymmetry breaking
$SU(3)\times SU(2)$ sector are 
\[ 
\begin{array}{c|cc}
& SU(3) & SU(2) \\ \hline
P & 3 & 2 \\
\overline{U} & \overline{3} & 1 \\
\overline{D} & \overline{3} & 1 \\
V & 3 & 1 \\
\overline{V} & \overline{3} & 1 \\
L & 1 & 2 \end{array},
\]
together with the tree-level superpotential for the 3-2 model 
$W_{3-2}=\lambda P\overline{U}L$. 

With this information at hand we can find out what the
effective F-term multiplying the messenger quarks $Q\overline{Q}$ 
is. For this we consider the effective theory below the mass
scales $M_S$ and $M_V$.  In order for this approximation to
make sense we need to assume that $M_V>\Lambda_{SU(3)}$, where
$\Lambda_{SU(3)}$ is the scale of the $SU(3)$ group of the 
3-2 model responsible for the breaking of supersymmetry. 
Below the scale $M_S$, the singlets $S,\overline{S}$ are integrated 
out resulting in the superpotential
\begin{displaymath} 
  W=-\frac{(V\overline{V})(Q\overline{Q})}{M_s}+
      M_V V\overline{V}+M_QQ\overline{Q}.
\end{displaymath}
It is useful to analyze this theory by treating the coefficient
of $V\overline{V}$ as an effective (field dependent) mass for $V\overline{V}$. This
effective mass is thus $M_V-\frac{Q\overline{Q}}{M_S}$. 
The effective scale of the $SU(3)$ is (after integrating 
out $V,\overline{V}$)  
\[ \Lambda_{eff}^7=(M_V-\frac{Q\overline{Q}}{M_S})
\Lambda_{SU(3)}^6.\]
Thus the effective superpotential is 
\begin{equation}
  W=\frac{M_V (1-\frac{Q\overline{Q}}{M_V M_S})\Lambda_{SU(3)}^6}
  {(P\overline{U})(P\overline{D})}+M_QQ\overline{Q}+\lambda P\overline{U}L.
\end{equation}
Provided that the VEV of the messenger quarks $Q,\overline{Q}$ 
vanishes, the dynamical superpotential breaks supersymmetry,
and an effective F-term is generated for the messenger
quarks. One can estimate the magnitude of this effective F-term 
to be 
\begin{equation}
  \label{Feff}
  F_{eff}=\frac{F\Phi^2}{M_SM_V},
\end{equation}
where $F$ is the magnitude of the F-terms in the supersymmetry
breaking sector, while $\Phi$ is the value of the VEV's in the
supersymmetry breaking sector. In the case of the 3-2 model
$\Phi \sim \Lambda_{SU(3)}$, $F\sim \Lambda_{SU(3)}^2$\footnote{We have 
suppressed the dependence on the Yukawa coupling $\lambda$. We assume 
throughout this paper that Yukawa couplings are ${\cal O}(1)$ and do
not affect the estimates of mass scales much.}.

The constraints on the parameters of all models of this sort are:

1. $\frac{F_{eff}}{M_Q}\simeq 10^{4.5}$ GeV to ensure the 
generation of the correct values of the gaugino and squark
masses~\cite{DNNS}.

2. $F_{max}\leq 10^{18}$ GeV$^2$, in order to suppress gravity
mediated soft supersymmetry breaking terms which would
reintroduce the problem of flavor changing neutral currents.

3. $M_Q\geq \frac{F_{eff}}{M_Q}\simeq 10^{4.5}$ GeV to avoid
the appearance of negative mass squared messenger quarks.

4. $M_V, M_S > \Lambda_{SU(3)}$ so that the presented 
effective theory approach is trustworthy. 

These four constraints will be the same in all composite 
intermediary models; the only difference is in the expressions
for $F_{eff}, M_Q,M_V$ and $M_S$ in terms of the dynamical
scales of the different strongly interacting gauge groups.

In the case of the 3-2 model discussed above, one has to use
Eq.~\ref{Feff} together with the expressions for $M_S,M_V$ and
$M_Q$ of Eq.~\ref{masses} obtained from confinement to get the bound on the 
``parameters'' $\Lambda_{SU(2)}, \Lambda_{SU(4)}$ and
$\Lambda_{SU(3)}$, where $\Lambda_{SU(2)}$ and $\Lambda_{SU(4)}$
are the scales of the confining groups and  $\Lambda_{SU(3)}$ is
the scale of the $SU(3)$ group of the 3-2 model. 
Putting these constraints together we obtain the bounds 
\begin{eqnarray}
  && \Lambda_{SU(4)}\geq 10^{15.75} \; {\rm GeV} \nonumber \\
  && 10^{15.75} \; {\rm GeV} \geq \Lambda_{SU(2)} \geq 
  10^{12.15} \; {\rm GeV} \nonumber \\
  && \Lambda_{SU(3)}\leq 10^{9} \; {\rm GeV},
\end{eqnarray}
such that the inequality $\Lambda_{SU(4)}^7 \Lambda_{SU(2)}^3
\leq 10^{157.5}$ GeV$^{10}$ is satisfied. The bound on 
$\Lambda_{SU(2)}$ is relaxed to 
$ 10^{17} \; {\rm GeV} \geq \Lambda_{SU(2)} \geq 
 10^{12.15} \; {\rm GeV} $ if we allow for a weaker $F_{max}\leq 10^{20}$
GeV$^2$ constraint instead of $F_{max}\leq 10^{18}$ GeV$^2$. 

In order to improve the allowed range of parameters of the
previous model one could consider 
 a different dynamical supersymmetry breaking
sector for the same kind of composite intermediary models. Clearly, if we
use the same model of compositeness we will not change the bound 
 $\Lambda_{SU(4)}\geq 10^{15.75} \; {\rm GeV}$ since this is purely a
consequence of the fact that the messenger quarks transform as
$5+\overline{5}$ under the ordinary $SU(5)$ group.
A different model of dynamical supersymmetry
breaking could lead to a better model, provided that one can take a
smaller representation than the $3+\overline{3}$ of the 3-2 model, since
in this case one would need to add a smaller confining group which would
result in weaker bounds on the confining scale. However, the 
3-2 model is one of the smallest models of dynamical supersymmetry breaking;
thus much improvement cannot be achieved on the above bounds either.

Let us summarize the field content and the interactions of the model
we introduced in this section. As a summary we repeat  the complete structure
of the model at high energies, above the compositeness scales at which the 
composite messenger quarks $Q$ and $\overline{Q}$ and the extra composite
flavor $V$, $\overline{V}$ are generated. The field content is given in
the table below.
\begin{displaymath}
  \begin{array}{c|ccccccc}
    &SU(2) & SU(3) & SU(3) & SU(2) & SU(4) & SU(5) & SU(5) \\ \hline
  P & 1   & 3     &   1   &   2   &  1    &   1   &   1   \\
  \overline{U}&1 &\overline{3}&   1   &   1   &  1    &   1   &   1   \\
  \overline{D}&1 &\overline{3}&   1   &   1   &  1    &   1   &   1   \\
  q & 2   & 3     &   1   &   1   &  1    &   1   &   1   \\
  \overline{q}& 2& 1     &\overline{3}&   1   &  1    &   1   &   1   \\     
  L & 1   & 1     &   1   &   2    &  1    &   1   &   1   \\  
  p & 1   & 1     &   1   &   1   &  4    &   5   &   1   \\
  \overline{p}&1 & 1     &   1   &   1   &  \overline{4}& 1   &\overline{5}  
  \end{array},
\end{displaymath}
where the first $SU(2)$ group is the confining $SU(2)$ producing the
composite $V$ and $\overline{V}$ fields. The two $SU(3)$ factors are 
the global symmetries of the confining $SU(2)$ group, where the diagonal
$SU(3)$ is gauged and is identified with the $SU(3)$ group of the
supersymmetry breaking 3-2 model. The second $SU(2)$ factor is the 
$SU(2)$ of the 3-2 model, and the $SU(4)$ is the confining group producing
the composite messengers. The two $SU(5)$'s are the global symmetries 
of the confining $SU(4)$ group, with the diagonal $SU(5)$ being identified 
with the ordinary $SU(5)$. The superpotential of this model is given at 
high energies by
\[ W= P\overline{U}L + \frac{1}{M_P^5} p^4\overline{p}^4 +
\frac{1}{M_P} q^2\overline{q}^2+ \frac{1}{M_P} q\overline{q}p\overline{p},\]
which then results in gauge mediation of supersymmetry breaking 
as described above. We emphasize again that this model does not have 
gauge singlet chiral superfields. However, the model is not completely chiral.
Therefore certain mass terms are forbidden by discrete
or global symmetries.

Finally, we comment on the relation of the composite
intermediary models and the models presented in Ref.~\cite{yael},
in which a  nonrenormalizable operator coupling the messengers and
the DSB sectors was assumed. In those models, it
was found that one generally required a new mass scale
to  suppress these operators; $M_P$ was too big.  So an operator
suppressed with an explicit new mass scale was necessary. It
seemed difficult to realize this composite dynamics, in which
both visible sector and DSB sector fields participated in common
gauge dynamics without introducing problematic new operators.
We therefore chose to realize this mass scale explicitly through
the mass of the $S,\overline{S}$ fields.  The dynamics of the two sectors
could then be separated and the operators which result are  
precisely those which are listed. However, the net result is similar
in spirit; a composite operator which links the two sectors
and obviates the need for a singlet F-term.

\subsection{Models without explicit mass for the messengers}

In this section we consider the possibility that there is no 
explicit mass term $M_Q$ present in the superpotential for the
messengers. 
To constrain the type of models we assume first that there
is a non-renormalizable tree-level superpotential of the form
$W=\frac{\Phi^k}{M_P^{k-3}}$ where $\Phi$'s are fields from the
supersymmetry breaking sector. We also assume that supersymmetry is
broken after the generation of a dynamical superpotential of the 
form $W_{dyn}=\frac{\Lambda^{p+3}}{\Phi^p}$. If we assume that there is
no explicit messenger mass $M_Q$ present, but the messenger mass comes 
purely from the expectation value of the effective operator of Eq.~\ref{effop}
then conditions 1-4 of the previous section imply $k\leq 5$, independent
of the details of the supersymmetry breaking sector, {\it i.e.}\ independent 
of $p$\footnote{A similar analysis shows that the constraints 1-4 are 
independent of the supersymmetry breaking sector ({\it i.e.}\
independent of p) even when an explicit mass term $M_Q$ for
the messengers is included. In 
this case the conditions 1-4 yield a constraint $(\frac{M}{M_P})^{k-3} M^2
M_Q<10^{31.5}$ ${\rm GeV}^3$ where the coupling between the
messengers $Q,\overline{Q}$ and the fields of the supersymmetry breaking
sector $\phi$ has been assumed to be of the form
$\frac{\phi^kQ\overline{Q}}{M^{k-1}}$.}. 
To allow a larger $\Lambda_{DSB}$ than $10^{9}$ GeV one needs to
consider theories with non-renormalizable operators.
Since $k\leq 5$ we consider a theory which has $k=4$, namely the
$SU(6)\times U(1)$ model of DNNS~\cite{DNNS}. 
In this theory the dynamical supersymmetry breaking sector is given by
\[ 
\begin{array}{c|cc}
  & SU(6)& U(1) \\ \hline
  A & 15 & 1  \\
  \overline{F}^+ & \overline{6} & -2 \\
  \overline{F}^- & \overline{6} &  -2\\
  S^0 & 1 & 3\\
  S^+ & 1 & 3\\
  S^- & 1 & 3\\
  V & 6 & 0 \\
  \overline{V} & \overline{6} & 0
\end{array},
\]
where we again added the extra vector-like flavor $V,\overline{V}$ to the 
field content. The non-renormalizable 
superpotential required for dynamical supersymmetry
breaking is
\[ W_{tree}=\frac{1}{M_{P}} A\overline{F}^+\overline{F}^-S^0.\] 
In order to generate the composite  $V,\overline{V}$ fields we need to 
introduce a confining $SU(5)$ group with six flavors, and the
resulting mass term for the $V$ is given by
\begin{equation}
\label{mv}
M_V=\frac{\Lambda_{SU(5)}^8}{M_P^7}.
\end{equation}
Similarly, the mass term for the singlets is given by
\[ M_S=\frac{\Lambda_{SU(4)}\Lambda_{SU(5)}}{M_P}.\]
The effective coupling between the messenger quarks and the
fields $\Phi$ of the supersymmetry breaking sector 
obtained after integrating out the fields $S,\overline{S},V,\overline{V}$ is
\[ \frac{\Phi^4}{M_PM_SM_V}Q\overline{Q}.\]
Since we have no explicit $M_Q$ mass term for the messenger quarks 
present, the effective $F/\Phi$ value just coincides with the
original value of $F/\Phi$ in the supersymmetry breaking sector,
which fixes the scale of the supersymmetry breaking $SU(6)$ group
to be $\Lambda_{SU(6)}=10^{9.9}$ GeV. Since supersymmetry breaking is
achieved through a non-renormalizable operator, the F-term in the
supersymmetry breaking sector is $F=\frac{\Lambda^{5/2}}{M_P^{\frac{1}{2}}}$.
The condition $F\leq 10^{18}$ GeV$^2$ results in the requirement
$\Lambda_{SU(6)}\leq 10^{10.8}$ GeV, which is clearly obeyed by the
required value of $\Lambda_{SU(6)}$. 

The constraint of having only positive mass-squared eigenvalues
in the messenger sector is different here than in Section~\ref{32sec}, 
since there is no
explicit mass term $M_Q$ present. Now the condition $M_Q^2>F_{eff}$ 
should be written as
\begin{displaymath}
 \left( \frac{\Phi^4}{M_P M_S M_V} \right)^2>\frac{F_{\Phi}\Phi^3}{M_PM_SM_V},
\end{displaymath}
which results in the constraint
\[ M_SM_V \leq 10^{22.5}\; \;  {\rm GeV}^2.\]
Thus the allowed range for the mass parameters $M_S,M_V$ is 
\[ 10^{10} \; {\rm GeV}\leq M_S,M_V \leq 10^{12} \; {\rm GeV},\]
such that $M_SM_V <10^{22.5}$ GeV$^2$. The $10^{10} \; {\rm GeV}\leq M_V$ 
constraint together with Eq.~\ref{mv} results in $\Lambda >10^{17}$ GeV.

Another possible model with a non-renormalizable term necessary 
for supersymmetry breaking is the $Sp(4)\times U(1)$ model of Ref.~\cite{Sp}.
The field content of the supersymmetry breaking sector is given by
\[
\begin{array}{c|cc}
  & Sp(4) & U(1) \\ \hline
  A & 5 & 2 \\
  q_1 & 4 & -3 \\
  q_2 & 4 & -1 \\
  s_1 & 1 & 2 \\
  s_2 & 1 & 4 \\
  V & 4 & 0 \\
  \overline{V} & 4 & 0
\end{array},
\]
where the field $A$ is in the traceless antisymmetric representation of 
$Sp(4)$ and we have again introduced the vector-like flavor $V,\overline{V}$ 
into the supersymmetry breaking sector.
The superpotential needed for supersymmetry breaking is
\[ W=q_1q_2s_2 + \frac{1}{M_P}q_1Aq_2 s_1.\]
The vacuum structure of this theory can be analyzed by first
assuming that the coefficient of the non-renormalizable operator
in the tree-level superpotential is turned off. Then there is a 
runaway direction, along which the $Sp(4)$ gauge group is completely 
broken. Turning the coefficient of the operator  $\frac{1}{M_P}q_1Aq_2 s_1$
on yields a finite vacuum in which $Sp(4)$ is completely
broken and the remaining uneaten singlets are massive. Carrying out this 
analysis with the requirement that the supersymmetry breaking F-term is less
than $10^{18}$ GeV$^2$ gives the bound $\Lambda_{Sp(4)}<10^{11.4}$ GeV. 
We again assume the same kind of compositeness as presented in 
Section~\ref{comp}, {\it i.e.} there is a confining $SU(3)$ group
producing the composite fields $V,\overline{V}$ and $S$, while the 
confining $SU(4)$ group produces the messenger quarks $Q,\overline{Q}$ and the
singlet $\overline{S}$. Since we assume no explicit $M_Q$ mass for the
messengers the condition $(F/\phi)_{eff}=10^{4.5}$ GeV fixes the $Sp(4)$ scale
to be $\Lambda_{Sp(4)}=10^{10.3}$ GeV, which is below the bound $10^{11.4}$
GeV obtained above. The remaining constraints yield the following
requirements for the scales
\[ \Lambda_{SU(3)}>10^{16} \; {\rm GeV}, \; \; 10^{14.5}\;{\rm GeV} >
\Lambda_{SU(4)} > 10^{10.3} \; {\rm GeV}.\]

\subsection{General considerations on variations of the 
composite intermediary models}

First, one should ask the question of why we
need the fields $V,\overline{V}$ at all. In this scenario one of the singlets
$S$ would be elementary and directly coupled to the supersymmetry 
breaking sector through a non-renormalizable operator, while the other
singlet $\overline{S}$ and the messenger quarks $Q,\overline{Q}$ are still
composite. One model would be for example to consider the 3-2 model
without an extra $SU(3)$ flavor, but with the additional coupling
\[ W=\frac{1}{M_P} SP\overline{U}L.\]
However the condition $F<10^{18}$ GeV$^2$ results in $\Lambda_{SU(4)}\leq
10^{14}$ GeV, which 
is incompatible with the
condition $\Lambda \geq 10^{15.75}$ GeV which comes from
$M_S>\Lambda_{SU(3)}$ and so this model is excluded. 
Since the most restrictive condition $\Lambda \geq 10^{15.75}$ GeV is a 
consequence of our choice of model of compositeness one could ask
whether it is possible to make this model work by choosing a 
different scheme for compositeness. For example one could instead of
gauging the diagonal $SU(5)$ subgroup of the $SU(5)\times SU(5)$
global symmetry of a confining  $SU(4)$ theory with five flavors we 
just gauge one of the global $SU(5)$'s and identify that with 
the ordinary $SU(5)$. In this case the field content of the confining
module is
\[ 
\begin{array}{c|c|cc}
& SU(4) & SU(5) & SU(5) \\ \hline
q & 4 & 5 & 1 \\
\overline{q} & \overline{4} & 1 & 5 \\
p_i & 1 & \overline{5} & 1 \\ 
\end{array},\ i=1,...,4.
\]
The fields $p_i$ are included to cancel the $SU(5)$ anomalies. The messenger
quark could be identified with one of the meson fields $q\overline{q}$ and the 
baryon $q^4$, while the singlet could be identified with a component of
the antibaryon $\overline{q}^4$. In this case the expressions for the 
composite masses for the singlets and the messenger quarks are 
modified to be
\[ M_S=\frac{\Lambda_{SU(4)}^3}{M_P^2}, \; 
M_Q=\frac{\Lambda_{SU(4)}^4}{M_P^3}. \]
Using these expressions together with the constraints 1-4 of 
Section~\ref{32sec} we get that the allowed range for $\Lambda_{SU(4)}$ is
\[ 10^{14.63} \; {\rm GeV} \leq \Lambda_{SU(4)} \leq 10^{14.78} \; {\rm GeV}.\]
The upper bound is extended to $10^{15.3}$ GeV if we allow for F-terms 
up to $10^{20}$ GeV$^2$. Thus we can see that there is no fundamental
need for the extra vector-like flavor $V+\overline{V}$ in the dynamical
supersymmetry breaking sector, but models
with vector fields have a larger allowed parameter range.

Next, we address the question of why we needed to rely in our composite
models on the dynamically generated confining superpotential terms. As
an alternative approach one could just add the appropriate tree-level
superpotential term for the preon fields by hand. In this case
however, the Yukawa coupling terms would not be of order one, 
but would be suppressed by powers
of the ratios of the confinement scales to the Planck scale. 
For example in the model with mass terms present for $V,\overline{V},S,\overline{S},Q,
\overline{Q}$ the effective $F/\phi$ is just
\begin{equation}
\label{f/phi} 
\lambda \frac{F^2}{M_SM_VM_Q}\sim 10^{4.5} \; {\rm GeV},
\end{equation}
where $\lambda = \frac{M_Q}{M_P}\frac{M_S}{M_P}
\frac{\Lambda_S^{p-1}}{M_P^{p-1}}$, where $\Lambda_S$ is the compositeness
scale of the group which binds $p$ preons into $S$. Thus
the constraint of Eq.~\ref{f/phi} is just
\[ \frac{F^2M_P^{p-3}}{M_V\Lambda_S^{p-1}}=10^{4.5} \; {\rm GeV}, \]
which together with $F\leq 10^{18}$ GeV$^2$ 
results in unacceptably small values of the product $M_V\Lambda_S^{p-1}$.
Thus we conclude that one really needs the order one Yukawa couplings 
resulting from confinement in order to obtain a viable spectrum
using the intermediary model.

Finally, we consider the possibility of using the  
dynamical supersymmetry breaking model of Ref.~\cite{su2} 
for the intermediary model. Recall 
that this model has an $SU(2)$ gauge symmetry with four doublets $q_i$
and six singlets $S^{ij}$ with a tree-level superpotential
$S^{ij}q_iq_j$. Since this model has naturally singlets and vector-like
flavors $q_i$ in it one would think that this model could be used as the
dynamical supersymmetry breaking sector in an intermediary model. This
is however not true. The reason is that the fields $q_i$ don't have an
F-term. This can be seen by writing down the scalar potential
\[ V=|{\rm Pf}M -\Lambda_{SU(2)}^4|^2+|M_{ij}|^2+
|S^{ij}+\lambda\epsilon^{ijkl}M_{kl}|^2,\]
where $\lambda$ is the Lagrange multiplier enforcing the quantum modified
constraint and $M_{ij}$ is the meson field $q_iq_j$. The point is that
the F-terms corresponding to $M_{ij}$ can always be satisfied by an 
appropriate choice of the singlet VEV's and thus the mesons will not
obtain F-term in this model of supersymmetry breaking, therefore one can
not use it directly in an intermediary model.

To summarize this section, we have shown two explicit composite intermediary
models which provide a viable sparticle spectrum similar to the
original DNNS spectrum. In one case, we used the 3-2 model as the 
dynamical supersymmetry breaking sector, and in the other case, we used 
the non-renormalizable $SU(6)\times U(1)$ model of DNNS for supersymmetry
breaking. Since all fields of these models are composites, there are 
no fundamental singlets present in these theories. The Yukawa couplings
are of order one due to confinement and the mass terms arise from tree-level
superpotential terms turning into masses after confinement. We have calculated
the viable parameter range for the confinement scales of these models.

\section{Mediator Models}

We begin by reviewing the  mediator models. 
There is a weakly gauged  messenger group   $G_m$ acting
on fields in the dynamical supersymmetry breaking sector.
This gauge group may or may not be broken, but a supersymmetry
breaking gaugino mass, $M_m$, for the gauge bosons of this group
is essential.  We assume this is achieved through a dynamical
messenger sector, such as those discussed in Refs.~\cite{pt,jmr}.
However, in the models of Ref.~\cite{pt,jmr} the weakly gauged group was 
$SU(5)$, so that supersymmetry breaking was communicated directly,
whereas we instead gauge a messenger gauge group $G_m$.
For definiteness, we
will take the messenger gauge group $G_m$ to be $SU(3)$.
This additional step in communicating
supersymmetry breaking solves the negative mass squared
problem for squarks and sleptons which was encountered in those
models~\cite{pt,jmr}.

Second, there 
are ``mediator fields", which we call $T$ and $\overline{T}$,    which transform
both under the messenger gauge group, $G_m$, and the standard
model gauge group, $G_{SM}$ (or an extension thereof)
and therefore ``talk" to both the dynamical supersymmetry
breaking and visible sectors.
Third, there is a supersymmetric mass term $M_T T \overline{T}$.  As discussed
in Ref. \cite{me},  $M_T$ is constrained to   lie between $ M_m$,
the messenger gaugino mass, 
and the mass of the heaviest of the  hidden sector scalars, $M_{DSB}$.  
 
In these models, there is no need to generate an $F$ term for a singlet
in order to generate the gaugino mass. Instead, it occurs from the three-loop diagram given in Figure 1.
\begin{figure}
\PSbox{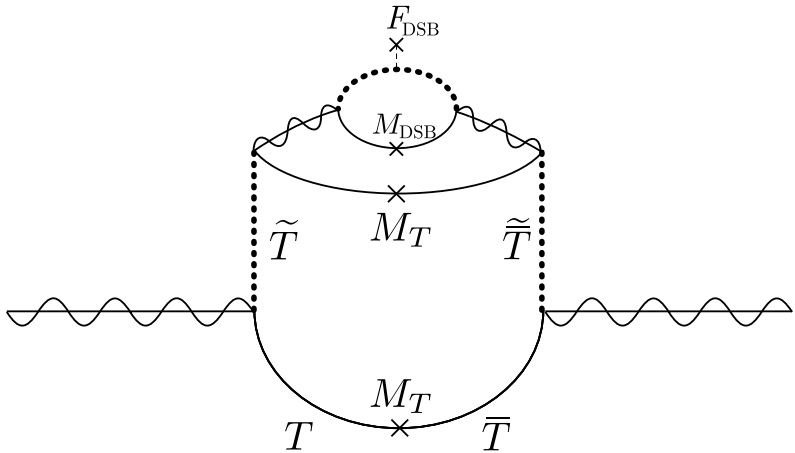 hoffset=10 voffset=-250 
hscale=70 vscale=70}{5cm}{6.5cm}
\caption{The diagram generating a visible sector gaugino mass in the 
mediator models of Ref.~\protect\cite{me}. $T$ and $\overline T$ denote
fermions, whereas $\widetilde T$ and $\widetilde{\overline T}$ denote scalars.}
\label{fig:F2}
\end{figure}

The scalar squared mass occurs at four-loops.  The leading
contribution can  readily
 be obtained  when there is  a large separation of mass scales as follows.
As computed in Ref. \cite{pt,jmr}, there is a negative logarithmically
enhanced
contribution to $S Tr M^2$ arising at two-loops. This feeds into
the squark and slepton mass squared with an additional two-loops.
The nice feature is the net contribution to the squark and slepton
mass squared is positive if the initial $S Tr M^2$ in the
messenger sector was positive, as it often is.  This
means that the model is phenomenologically viable, although
it can be more fine-tuned than the more conventional gauge-mediated
mass predictions. In particular, there is a relatively large 
prediction for the scalar to gaugino mass ratio (of like charges)
which is \footnote{ We thank Asad Naqvi for sharing his result.}
\beq
{m_i \over m_{i,1/2}}\approx \alpha_m {50 \over N_f} \left({k_i 
\over k_m}\right)^{1/2}
\left[{\beta Log \left({M \over m_f}\right)^2-N_f \over 
Log\left({M\over M_T}\right)^2 } \right]^{1/2}>1
\eeq
where $N_f$ is the number of flavors charged under $G_m$
and can be order 10, $\beta$ is a number of ${\cal O} (1)$, $M$ is the
biggest mass scale in the supersymmetry breaking sector, 
$\alpha_m=g_m^2/4\pi$, where $g_m$ is the messenger gauge coupling. 
Therefore the ratio, although large, might be viable.

One interesting feature of these models is that they
have the correct global minimum. That is, the minimum
is the color and charge-preserving minimum in which
supersymmetry is broken. This model is the only
known example we know with this property (in which
additional fields are not added ad hoc in order to guarantee
this property).

We now address the question of how to realize these
models in such a way that the mass term is not introduced
by hand. As we will see, this is fairly straightforward.

\subsection{A Composite Mediator Model}

In the explicit model we present the mediator 
fields $T,\overline{T}$ are composites 
of a confining $Sp(4)\times Sp(4)'$ gauge group. Each of the $Sp(4)$ 
groups has eight fields transforming as a fundamental representation $4$.
A gauged $SU(5)$ subgroup of the global $SU(8)$ symmetry is identified
with the ordinary $SU(5)$ group and a gauged $SU(3)$ group is identified
with the messenger $SU(3)_m$ gauge group. The fields transforming under the
confining $Sp(4)$ groups are summarized in the table below\footnote{Note 
that similar product group structures involving symplectic groups arise 
naturally in the orbifold construction of Ref.~\cite{Ken}.}.
\[
\begin{array}{c|cccc}
&SU(5)_{SM} & SU(3)_m &Sp(4) & Sp(4)' \\ \hline
T_5 & 5 & 1&4 &1 \\
T_3 & 1& 3& 4 &1 \\
\overline{T}_{\bar{5}} & \overline{5} & 1 & 1 & 4 \\
\overline{T}_{\bar{3}} & 1 & \overline{3} & 1 & 4 \\
\end{array}
\]
The bound state spectrum of this theory is given by
\[ \begin{array}{c|cc}
& SU(5) & SU(3)_m \\ \hline
\big (T_5^2\big) & 10 & 1 \\
\big(T_3^2 \big) &  1 & \overline{3}  \\ 
\big(\overline{T}_{\bar{5}}^2\big ) & \overline{10} & 1 \\ 
\big (T_{\bar{3}}^2\big) &  1 & 3 \\
T\equiv \big(T_5 T_3\big) &  5 & 3 \\ 
\overline{T} \equiv\big(\overline{T}_{\bar{5}}
\overline{T}_{\bar{3}}\big) &  \overline{5} &  \overline{3} 
\end{array}. \]
Thus the bound states $\big(T_5 T_3\big)$ and $\big(\overline{T_{\bar{5}}}
\overline{T_{\bar{3}}}\big) $ are 
identified with the mediator fields $T,\overline{T}$.
In order to generate mass terms for the composite mediator fields we have to
assume that a tree-level superpotential
\begin{displaymath}
  W_{tree}= \lambda_1T_5^2\overline{T}_{\bar{5}}^2+
  \lambda_2  T_5 T_3\overline{T}_{\bar{5}} \overline{T}_{\bar{3}}+
  \lambda_3 T_3^2\overline{T}_{\bar{3}}^2
\end{displaymath}
is present in the theory for the preon fields. 
After confinement, the tree-level Yukawa coupling are turned into mass terms
and a confining superpotential ${\rm Pf} M$ is generated 
resulting in the low energy superpotential
\begin{displaymath}
  W={\rm Pf} M +{\rm Pf} M'+\lambda_1(T_5^2)(\overline{T}_{\bar{5}}^2)+
  \lambda_2  T \overline{T}+
  \lambda_3 (T_3^2)(\overline{T}_{\bar{3}}^2).
\end{displaymath}
Thus, the $T\overline{T}$ mass term is given by 
$M_T=\Lambda_{Sp(4)}\Lambda_{Sp(4)'}/M_{P}$. Assuming the two $Sp(4)$ 
groups have scales of the same order, and assuming that the masses in the dynamical 
supersymmetry breaking sector lie between $10^4$ GeV and $10^9$ GeV, we get a bound 
on the confining scale of $10^{11}$ GeV $<\Lambda_{Sp(4)}<10^{13.5}$ GeV.
This model, besides being quite simple also has the advantage that the
F-term in the supersymmetry breaking sector is not required to be
close to $10^9$ GeV. Therefore this model can also satisfy the
constraints of Ref.~\cite{mur} coming from nucleosynthesis.

\section{Conclusions}

We have presented explicit realizations of the ``Intermediary'' and 
``Mediator''
models of gauge mediated supersymmetry breaking
presented in Ref.~\cite{me}. Both of these models have non-vanishing
tree-level mass terms as essential ingredients. In the models we presented 
these mass terms arise due to confining dynamics. The mass scales are not put
in by hand but rather determined as a function of the dynamical scales of the 
confining groups and the Planck scale. These masses always arise after
tree-level Yukawa couplings (sometimes non-renormalizable operators
suppressed by the Planck scale) turn into mass terms for the confined
low-energy degrees of freedom. The intermediary models also require order
one Yukawa couplings of the {\it confined} fields. In our models, these are
generated dynamically via the confining superpotentials of
Ref.~\cite{SeibergPRD}. 

The intermediary models have two singlets but no messenger gauge groups. The 
interactions of these singlets with the messenger quarks and with the
fields in the dynamical supersymmetry breaking sector generate an effective
operator that has similar effects as the singlet coupling to the messenger
quarks in the DNNS models. In the composite versions of this model both
singlets and the messenger quarks are composite and thus there is no
elementary singlet present at high energies. We have presented several
models of this sort and given the allowed range of the parameters of the
theory. The mediator models employ two fields $T,\overline{T}$ which carry
both the charges of the messenger gauge group and of ordinary $SU(5)$.
However, an explicit mass term for these fields together with masses
for the messenger gluinos are crucial for this
model to work. We presented an example in which the mass term of the mediators
arises via confinement where tree-level Yukawa terms turn into masses for the
composite mediator fields.

The models presented here, while being explicit realizations of the scenarios 
presented in Ref.~\cite{me}, are new complete examples of gauge mediation
of dynamical supersymmetry breaking.  The intermediary models don't have a
messenger gauge group and the composite versions of these models do not
contain elementary gauge singlet fields either, thus presenting a 
simplification compared to the original model of Ref.~\cite{DNNS}.
The mediator models do have a messenger gauge group, but the additional
structure required for gauge mediation is simpler than in the 
intermediary models. The explicit models presented in this paper 
represent viable alternatives to the conventional models of gauge mediated
supersymmetry breaking.

\end{document}